\documentclass[aps,prl,showpacs,twocolumn,groupedaddress]{revtex4}
\usepackage{graphicx}

\begin{document}


\title{Collision-Dependent Atom Tunneling Rate in Bose-Einstein Condensates}


\author{B. R. da Cunha and M. C. de Oliveira}
\affiliation{ Instituto de F\'\i sica Gleb Wataghin,  Universidade
Estadual de Campinas, 13083-970, Campinas - SP, Brazil.}


\date{\today}

\begin{abstract}
We show that the interaction (cross-collision) between atoms
trapped in distinct sites of a double-well potential can
significantly increase the atom tunneling rate for special trap
configurations leading to an effective linear Rabi regime of
population oscillation between the trap wells. The inclusion of
cross-collisional effects significantly extends the validity of
the two-mode model approach allowing it to be alternatively
employed to explain the recently observed increase of tunneling
rates due to nonlinear interactions.
\end{abstract}
\pacs{03.75.Lm,  74.50.+r}

\maketitle



 Atom optics and the physics of ultracold matter waves have
witnessed rapid theoretical and experimental progress since the
achievement of atomic vapor Bose-Einstein condensates (BECs)
\cite{BEC}. The interest in such a system is quite wide ranged as
it has opened new technological frontiers 
\cite{zoller,bectelep}, 
  and has renewed the investigation on many-body physics fundamental issues once
  BECs trapped in optical lattices offers a powerful
toolbox for implementation of condensed matter model systems with
highly controllable parameters \cite{leggett,bloch,qtbox}.

The simplest system to exhibit non-trivial many-body phenomena is
the two-mode BEC trapped in a double well potential.
 For this system it was previously predicted
\cite{corney1,raghavan1} a {\it plethora} of dynamical regimes,
including Josephson oscillation \cite{javanainen} of atoms between
distinct trap sites, and macroscopic self-trapping of atomic
populations, where coherent tunneling is suppressed when the
number of trapped atoms exceeds a critical value. The first
realization of an atomic single BEC Josephson junction has indeed
experimentally demonstrated \cite{albiez} those predictions
\cite{javanainen,corney1,raghavan1}. Moreover an increase of one
order of magnitude of the atomic tunneling rate (from 2 to 25 Hz)
due to nonlinear atomic interactions was then observed, which
cannot be explained inside those previous models.
In  those models \cite{corney1,raghavan1} a two-mode approximation
(TMA) is invariably applied, in which each trap well is populated
by one single localized condensate mode. However due to a
localization assumption the BEC dynamics is considered by
regarding only self-collisions (on-site interaction) between atoms
of individual condensate modes. Notwithstanding, distinct trap
configurations can be as such that cross-collisions (off-site
interaction) of atoms in distinct BEC modes cannot be neglected
and the observed coherent phenomena are affected by
cross-collision induced effects. This situation has been
previously considered either by directly integrating a
non-polynomial Schr\"odinger equation \cite{albiez,salasnich} or
by extending the TMA through Gross-Pitaeviskii equations
\cite{ananikian}.


In this paper we investigate the cross-collisional effects over
the quantum dynamics of a BEC trapped in a double-well potential
inside the TMA. By appropriate manipulation of the trapping
potential the system shows a very rich quantum dynamics related to
the cross-collisional rates.
 In absence of cross-collisions, self-trapping
is observed in agreement with previous theoretical predictions
\cite{corney1,raghavan1}. Remarkably, the inclusion of
cross-collisional effects can inhibit self-trapping, resulting in
an effective coherent oscillation (effective Rabi regime) of atoms
between the two wells with an effective Rabi frequency, dependent
on the cross-collision rate and the number of atoms in the sample.
Moreover cross-collisions increase the range of validity  of the
TMA (when the many-body interactions produce only slight changes
on the ground-state properties of the individual potential sites).
The new limit  can be as such that the TMA is indeed valid for the
number of atoms present in the experiment of Ref. \cite{albiez},
and thus can be employed for alternative explanation
\cite{albiez,ananikian} of the observed increase of tunelling rate
through off-site interactions.

We consider an atomic BEC trapped in a symmetric double-well
potential $V({\bf r})$ with minima at ${\bf r_1}$, and ${\bf
r_2}$, such that $V({\bf r_{1,2}})=0$, the many-body Hamiltonian
is given by
\begin{eqnarray}
\hat{H} &=&\int d^{3}r\hat{\psi} ^{\dagger }({\bf r})\left(-\frac{\hbar^2 }{2m}\nabla^{2}+V({\bf r})\right) \hat{\psi} ({\bf r})  \nonumber \\
&&+\frac{1}{2}U_0\int d^{3}r\hat{\psi} ^{\dagger }({\bf %
r})\hat{\psi} ^{\dagger }({\bf r})\hat{\psi} ({\bf r})\hat{\psi}
({\bf r}), \label{bec}
\end{eqnarray}
where $m$ is the atomic mass, $U_0=4 \pi\hbar^2a/m$ measures the
strength of the two-body interaction, $a$ is the s-wave scattering
length, $\hat{\psi}({\bf r}, t )$ and $\hat{\psi}^\dagger({\bf r},
t)$ are the Heisenberg picture field operators. We must emphasize
that throughout our calculations we assume that the barrier height
separating the two wells is sufficiently larger than the
condensate chemical potential, $ V_{0}>\mu_{c}$, so that a TMA can
be considered to describe the system \cite{williams}. The
derivation of the two-mode Hamiltonian in a double-well trap
follows the standard procedures of Ref. \cite{corney1} - The lower
energy eigenstates of the global double-well are approximated as
the symmetric and anti-symmetric combinations $ u_{\pm}({\bf
r})\approx\frac{1}{\sqrt{2}}[u_{1}({\bf r})\pm u_{2}({\bf r})]$,
and the field operators are expanded in terms of the local modes
$u_j(r)$, $j=1,2$ and the Heisenberg picture annihilation operator
as $\hat{a}_j(t)=\int d^3{\bf r} u^*_j({\bf r})\hat{\psi}({\bf
r},t)$ so that $[\hat{a}_j , \hat{a}_k^\dagger]\simeq\delta_{jk}$.
The potential expanded around each minimum is $V({\bf
r})=\widetilde V^{(2)}({\bf r}-{\bf r}_j)+...\;\;$j=1,2, where
$\widetilde V^{(2)}({\bf r}-{\bf r}_j)$ is the harmonic
approximation to the potential in the vicinity of each minimum.
The normalized single-particle ground-state $u_j({\bf r})$ of the
local potential $\widetilde V^{(2)}({\bf r}-{\bf r}_j)$, with
energy $E_0$, defines the local mode solutions of the individual
wells. The tunneling frequency $\Omega$ between the two minima is
then given by the energy level splitting of these two lowest
states, $\Omega=2R/\hbar$, where $R=\int d^{3}{\bf
r}u^{*}_{1}({\bf r})[V(r)-\widetilde{V}^{(2)}({\bf r}-{\bf
r}_{1})]u_{2}({\bf r}).$ By noticing that the total number of
atoms $N$ is a conserved quantity, after some algebraic
manipulation, the following Hamiltonian is then obtained in the
interaction picture
\begin{eqnarray}\label{hamilt}
\hat{H}&=&\hbar[2\Lambda(N-1)+\Omega](\hat{a}_1^{\dag}\hat{a}_2+\hat{a}_2^{\dag}\hat{a}_1)+\hbar\eta(\hat{a}_1^{\dag}\hat{a}_2+\hat{a}_2^{\dag}\hat{a}_1)^{2}\nonumber\\
&&+\hbar(\kappa-\eta)
[(\hat{a}_1^{\dag})^{2}(\hat{a}_1^{2})+(\hat{a}_2^{\dag})^{2}(\hat{a}_2^{2})]
+\hbar\eta N(N-2)
\end{eqnarray}
where $\kappa=\frac{U_{0}}{2\hbar V_{eff}}$ is the self-collision
rate and $\eta=(\frac{U_{0}}{2\hbar})\int d^{3}{\bf
r}u_{i}^{*}u_{j}u_{i}^{*}u_{0j}$, and
$\Lambda=(\frac{U_{0}}{2\hbar})\int d^{3}{\bf
r}u_{j}^{*}u_{j}u_{i}^{*}u_{i},$ are cross-collisional rates. Here
$V_{eff}^{-1}=\int d^{3}{\bf r}|u_{j}|^{4}$ is the $j$-mode
effective volume. It is immediate that for $\hbar\kappa\rightarrow
0$, the Hamiltonian (\ref{hamilt}) reduces to that previously
investigated in the study of tunneling in condensates
\cite{javanainen}. Also for $\hbar\eta\rightarrow 0$ and
$\hbar\Lambda\rightarrow 0$, the Hamiltonian (\ref{hamilt})
reduces itself to that discussed in \cite{raghavan1,corney1},
accounting for tunneling oscillations as well as for population
self-trapping. As $\eta\rightarrow\kappa$ in Hamiltonian
(\ref{hamilt}) a new dynamical regime of stable long-time
tunneling (hereafter called effective Rabi regime) can be
attained.


Introducing the Schwinger angular momentum operators
$\hat{J_{x}}=(\hat{a}_2^{\dag}\hat{a}_2-\hat{a}_1^{\dag}\hat{a}_1)/2,
\hat{J_{y}}=i(\hat{a}_2^{\dag}\hat{a}_1-\hat{a}_1^{\dag}\hat{a}_2)/2,
\hat{J_{z}}=(\hat{a}_1^{\dag}\hat{a}_2+\hat{a}_2^{\dag}\hat{a}_1)/2,$
the Hamiltonian (\ref{hamilt}) can be rewritten as
\begin{equation}\label{hamilt2}
\hat{H}=\hbar\Omega'\hat{J}_{z}+4\hbar\eta\hat{J}_{z}^{2}+2\hbar(\kappa-\eta)\hat{J}_{x}^{2},
\end{equation}
where  we have neglected constant energy shifts dependent on $N$
and have defined a new effective tunneling rate $\Omega'\equiv
2\left[\Omega+2\Lambda(N-1)\right]$, which is dependent on the
number of atoms in the atomic sample and the cross-collisional
rate $\Lambda$. In fact $\Lambda$ is very small compared to
$\Omega$, but for a sufficiently large number of atoms the
additional tunneling term may lead to observable effects whenever
$2\Lambda (N-1)\geq\Omega$. Furthermore the third term of Eq.
(\ref{hamilt2}) shows that the cross-collisional rate $\eta$
competes with the self-collision rate $\kappa$ leading to an
effective on-site collision rate $\kappa^\prime\equiv
\kappa-\eta$. Since $0\le\eta\le\kappa$, $\kappa^\prime$ on Eq.
(\ref{hamilt2}) can be disregarded in a trap configuration where
$\eta\rightarrow\kappa$ with $\Omega'\gg\kappa'$. To infer the
validity and implications  of such a regime we consider a specific
trapping potential \cite{footnote2} of the form $ V({
r})=b\left(x^{2}-\frac{d}{2b}\right)^{2}+\frac{1}{2}m\omega^{2}_{t}(y^{2}+z^{2})$,
where the inter-well coupling occurs along $x$, and $\omega_{t}$
is the trap frequency in the y-z plane. This potential has fixed
points at $q_{o}^2=\frac{d}{2b}$. The position uncertainty for a
harmonic oscillator in the ground state is
$x_{0}\equiv
\sqrt{\frac{\hbar}{2m\omega_0}}$, with $\omega_0=\sqrt{4d/m}$. For
a suitable choice of the barrier height only two energy
eigenstates lie beneath the barrier. Assuming $\omega_t=\omega_0$
the local mode on each well is then given by
$u_{j}(r)=(\frac{1}{2\pi
x_{0}^{2}})^{\frac{3}{4}}\exp\left(\frac{-[(x_{j}-q_0)^2+y_{j}^2+z_{j}^2}{4x_{0}^{2}}\right).$
 The
collision rates may then be evaluated to give 
$\kappa =(\frac{U_{0}}{16\hbar})(\frac{1}{\pi x_{0}^{2}})^{3/2}$,
 $\eta=\kappa \exp(-q^{2}_{0}/2x_{0}^{2})$, and $\Lambda=\kappa
\exp(-3q^{2}_{0}/4x_{0}^{2})$, while
$\Omega=\frac{q_0^2\omega_0}{x_0^2}\exp{(-q^{2}_{0}/2x_{0}^{2})}$.

Firstly we shall investigate the validity of the localization of
the TMA considered in the present description. It is valid only if
the many-body interactions produce small modifications on the
ground-state of the individual potential wells \cite{footnote3}.
In the absence of cross-collisions \cite{corney1} the TMA is
limited to $ \hbar\omega_{0}=\frac{\hbar^{2}}{2mx_{0}^{2}}\gg
N\hbar\kappa= N\frac{\hbar|U_{0}|}{2\hbar V_{eff}}$ and thus the
BEC is limited to a few 100 atoms. The inclusion of
cross-collision necessarily increases the effective mode volume,
which in terms of the effective on-site collision rate $\kappa'$
is given by $V_{eff}'\equiv V_{eff}/(1-e^{-q_0^2/2x_0^2})$. For
the trapping potential considered, $V_{eff}\sim
8\pi^{3/2}x_{0}^{3}$, and
%
the TMA is valid whenever
\begin{equation}
N\ll\frac{2\pi^{1/2}x_{0}}{|a|(1-e^{-q_{0}^{2}/2x_{0}^{2}})}.
\end{equation}
For suitable values of the trap parameters, it is thus possible to
have very large numbers at the right of the inequality above.
Hence the considered TMA, with cross-collision included, can be
appropriately employed for the description of the observed
phenomena in present experimental setups
\cite{albiez} for numbers of atoms $N\gtrsim 10^{3}$. 

Typically $\eta$, $\Lambda\ll \Omega$, but 
cross-collisional effects can only be neglected if
$2\Lambda(N-1)/\Omega\ll 1$, {\it i. e.} if $(N-1) \frac{a_s
x_0}{\pi^{1/2}q_0^2}e^{-q_0^2/4x_0^2}\ll 1$. By considering the
experimental values from Ref. \cite{albiez} for $N=1300$,
$a_s=5.3$ nm for $^{87}$Rb atoms, and $q_0=2.2 \mu$m, we obtain
that only for $x_0\ll 0.82\; q_0$ is that cross-collisional
effects can be neglected. This bound value however depends
strongly on the number of atoms in the sample and decreases
rapidly as $N$ is increased. For example, if the number of atoms
is increased in one order of magnitude to $N=13000$ then the
cross-collisional effects can be neglected only if $x_0\ll 0.366\;
q_0$. Furthermore by equating $\Omega^\prime$ to the observed
frequency of oscillation of  25 Hz in the experiment performed by
Albiez {\it et al.} \cite{albiez} and $\Omega$ to the expected
frequency of tunneling of 2 Hz we find with the above parameters
that only for $x_0\approx 6.56\; q_0$ is that the tunneling
frequency would significantly be increased due to nonlinear
off-site interactions. Of course, inside the applied TMA this
situation corresponds to that of two strongly overlapped modes,
which would compromise the validity of the localization assumption
and the considered gaussian eigenfunctions. However the relation
between $x_0$ and $q_0$ is not linear and depends strongly on $N$
and on $q_0$ as well. For a fixed $q_0$ the $x_0$ to $q_0$ ratio
decreases rapidly as $N$ is increased. Again, for $N=13000$ the
substantial increase in the tunneling frequency would be observed
for $x_0\approx 0.89\; q_0$. We remark that the above analysis
could be similarly applied by considering the local modes as
approximately given by the excited states of the harmonic
oscillator. In such a case the localization conditions in relation
to the distance between minima of the trapping potential would be
significantly relaxed \cite{dassarma}.
\begin{figure}[h]
\centering\vspace{-0.5cm}
\includegraphics[width=3.35in]{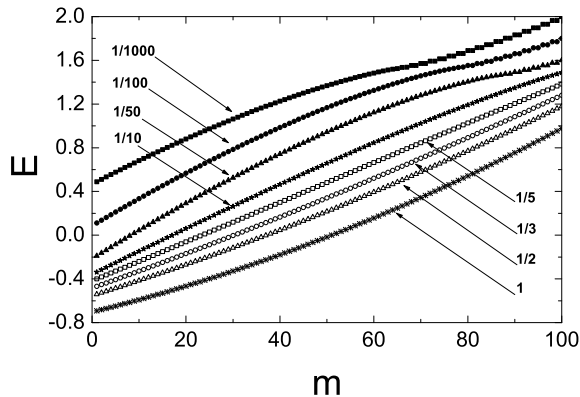}\hspace*{5cm}\vspace*{-0.8cm}
\caption{Numerical results for the eigenvalues of Eq. 3 in the
eigenbasis of $\hat{J}_{x}$. The energy spectra is shown for
different regimes of the $\eta/\kappa$ ratio. Each spectrum has
been displaced by a constant factor.}
\end{figure}

We now turn to the description of the dynamical regimes attained
by the Hamiltonian (\ref{hamilt2}).
 In Fig. 1 we plot the numerically calculated eigenvalues spectra for a
condensate with 1000 atoms and fixed $\Omega/\kappa=50$ ratio.
Each curve in the plot represents a specific $\eta/\kappa$ ratio
as depicted in the figure. For $\eta/\kappa=1/1000$ it is
immediately verified that the eigenvalues correspond to those of
the model without cross-collision discussed in Ref.
\cite{corney1}. The low-excited eigenstates are close to the
eigenstates of the $\hat{J}_z$ operator given by $|j,-j\rangle_z$.
Highly excited eigenstates are closer to the eigenstates of the
operator $\hat{J}_x^2$, given approximately by $|j,0\rangle_x$.
This last state suggests an ideal situation for occurrence of
population self-trapping for large $N$. The inflexion point for
intermediate eigenvalues represents the frontier for the
occurrence of self-trapping. As the ratio $\kappa/\eta$ is
increased the eigenvalues behavior changes dramatically due to the
influence of the term proportional to $\hat{J}_z^2$. An
outstanding feature is that the inflexion point is shifted to
higher eigenvalues for $\kappa/\eta=1/100 - 1/10$ and completely
disappears for $\kappa/\eta\ge 1/5$. For $\kappa/\eta\approx 1$
the term proportional to $\hat{J}_x^2$ is negligible in comparison
to the other terms and the Hamiltonian eigenstates get as closer
to the $\hat{J}_z$ eigenstates as $\kappa/\eta\rightarrow 1$.

The equations of motion for the angular momentum operators are
given by
\begin{eqnarray}\label{h1}
\dot{\hat{J}_{x}}&=&-\hbar\Omega^\prime\hat{J}_{y}-4\hbar\eta[\hat{J}_{y},\hat{J}_{z}]_{+}\\
\label{h2}
\dot{\hat{J}_{y}}&=&\hbar\Omega^\prime\hat{J}_{x}-2\hbar(\kappa-3\eta)[\hat{J}_{z},\hat{J}_{x}]_{+}\\
\label{h3}
\dot{\hat{J}_{z}}&=&2\hbar(\kappa-\eta)[\hat{J}_{y},\hat{J}_{x}]_{+}
\end{eqnarray}
One can immediately see that if the cross-collision term is strong
enough so that $\kappa-\eta\ll\Omega'$ then
\begin{equation}\label{h4}
\dot{\hat{J}_{z}}\approx0
\end{equation}
and thus $\hat{J}_{z}$ being a constant of motion the set of Eqs.
(\ref{h1}, \ref{h2}, and \ref{h4}) is solved for $\hat{J}_{x}$ as
\begin{equation}
\hat{J}_{x}(t)=\hat{J}_{x}(0)\cos\Omega^"t+\hat{J}_{y}(0)\sin\Omega^"t,
\end{equation}
where $\Omega^"\equiv\Omega^\prime+8\eta
J_z=\left[\Omega+2\Lambda(N-1)\right]+8\eta J_z$ is the new Rabi
frequency, which now depends on both the number of atoms and the
initial condition for $J_z$ through the nonlinear cross-collision
rates $\Lambda$ and $\eta$, respectively. Since $\eta\ll\Omega$,
$J_z$ value only gives a small shift in $\Omega^"$. The effective
Rabi regime of $\eta\rightarrow\kappa$ is justified whenever
$\kappa-\eta\ll\Omega''$.


In Fig. 2 the time evolution of the mean value
$\langle\hat{J}_{x}\rangle$ as given by the general system of Eqs.
(5-7) for distinct $\eta/\kappa$ ratio. We suppose an initial
state localized in one well as $|j,-j\rangle_{x}$. Similarly to
Ref. \cite{corney1} due to intrinsic quantum fluctuations in the
initial condition there are some oscillations of the quantum mean
decay \cite{andersonb}. In Fig. 2(a) the revival of the
oscillation that occurs at later times is again due to the
discrete spectrum of the many-body Hamiltonian
\cite{bernstein,chefles}. In Fig. 2(b) the self-trapping of the
atomic population is shown. This result should be compared to that
of Fig. 2 from Ref.\cite{albiez}. More interesting though are the
new features in the quantum dynamics due entirely to the presence
of cross-collisions appearing in Fig. 2(c). Notice that even when
cross-collisions are relatively small ($\kappa/\eta\sim10$)
coherent tunneling dominates, suppressing the initial
self-trapping. As the cross-collisions become stronger the
collapse and revival dynamics changes. For large intervals of time
there is a total collapse of the mean value, which lasts for large
periods of time in such a way that this period decreases as $\eta$
gets closer to $\kappa$ approaching the limit where it tends to
zero when the effective Rabi regime takes place and there is no
collapse and revival. Notice that the decreasing amplitude
modulation in the collapse and revival is due to nonlinearities
introduced in Eqs. (5-7) by cross-collisions. As
$\eta\rightarrow\kappa$ however, the amplitude of coherent
oscillation of the population difference goes to a constant value
with oscillating frequency $\Omega^"$.

To push the discussion presented here further, it is interesting
to consider the impact of the cross-collisional terms in related
many wells systems \cite{qtbox}. The dynamics is richer and
additional effects may occur as consequence of the overlap of
neighboring condensed modes. The results found for the double-well
case can be extended and set into the more general framework of
hopping in many wells systems. It is possible to render a general
discussion of the cross-collisional terms for the derivation of a
discrete extended Bose-Hubbard model \cite{dassarma} and of the
Wanier-Stark effect \cite{buchleitner} based solely on ground
state trap solutions. For instance, a $n$-mode {\it ansatz} of the
natural extension of the Hamiltonian in Eq.(2) results in a
$n$-mode extended Bose-Hubbard Hamiltonian
\begin{eqnarray} \label{BH}
\hat{H}&=&\sum_{l}(\kappa_{l}-\eta_{l,l+1})\hat{n}_{l}(\hat{n}_{l}-1)\nonumber\\
&&+\sum_{l}\eta_{l,l+1}(\hat{a}^{\dag}_{l+1}\hat{a}_{l}+H.c)^{2}\\
&&+\sum_{l}[\Omega_{l,l+1}+2\Lambda_{l,l+1}(2\hat{n}_{l}-1)](\hat{a}^{\dag}_{l+1}\hat{a}_{l}+H.c)\nonumber,
\end{eqnarray}
with all parameters related to the those for the double-well
system. The dynamical regimes given by (\ref{BH}) have in part a
similar aspect to the ones presented for the double-well potential
but can be extended since all the on-site and off-site parameters
can be independently varied. Those issues are being further
investigated in the context of phase transitions and shall be
addressed elsewhere \cite{bhm}.
\begin{figure}[h]
\centering\vspace{-.2cm}
\includegraphics[width=3.15in]{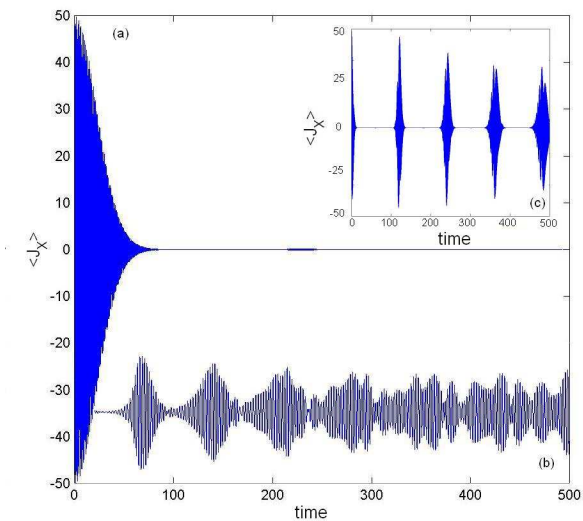}\vspace*{-4.3cm}
\caption{Numerical results for the mean value of the population
imbalance $\langle\hat{J}_{x}\rangle$ for N=100, $\kappa N=2.0$,
(a) $\eta=\kappa/10$, (b) $\eta=\kappa/100$ and
(c)$\eta=\kappa/2$. For convenience we normalize the time in units
of the single particle tunneling period so that in these units
$\Omega=1$. Time evolution found by integrating Schr\"odinger
equation in the $\hat{J}_{x}$ basis for the Hamiltonian in Eq. 3.}
\end{figure}

In summary we employed the TMA to model an atomic two-mode BEC
trapped in a double well potential considering cross-collisions
between the condensed modes. Cross-collisions strongly inhibit the
self-trapping phenomenon even for small values of the
cross-collision rate $\eta$. For a given trap geometry the
eigenvalues of the many-body Hamiltonian determine a new dynamics
resulting in an effective coherent oscillation (effective Rabi
regime) of the population in both wells. In the limit
$\eta\rightarrow\kappa$ for $\kappa-\eta\ll\Omega^"$ an effective
Rabi regime takes place with an effective Rabi frequency,
$\Omega^"$, which is explicitly dependent on the total number of
atoms. The limit of validity of such a two-mode approximation
increases within the assumption of strong cross-collisions in such
a way that it may be in accordance with experimental data
\cite{albiez} observed for the increase of tunneling frequency due
to nonlinear interaction. As a last result the collapse and
revival dynamics is changed resulting in large periods of total
collapse. As $\eta\rightarrow\kappa$, however the collapse and
revival frequency increases attaining the regime of Rabi
population oscillation for $\eta\sim\kappa$. It is remarkable that
similar results to those previously given in exact numerical
calculations \cite{albiez,salasnich,ananikian} could be derived
inside the TMA just by adding cross-collisional effects.
We hope that the above considerations bring some contribution to
experimental investigation.

The authors would like to acknowledge partial financial support from
FAPESP under project $\#04/14605-2$, from CNPq and from
FAEPEX-UNICAMP.


\begin{thebibliography}{99}
\bibitem{footnote2} We remark that the results obtained are qualitatively
equivalent for any double-well trapping potential.
\bibitem{footnote3}
Otherwise the Thomas-Fermi approximation \cite{baym,edwards}
should be applyied  instead of the TMA.
\end{thebibliography}
\end{document}